\documentstyle[aps,epsfig]{revtex}

\begin{document}
\def\be{\begin{equation}}
\def\ee{\end{equation}}

\title{Memory in Self Organized Criticality}

\author{Eugenio Lippiello,$^1$ Lucilla de Arcangelis$^2$ and
Cataldo Godano$^3$}
\address{ $^1$ INFM, UdR Naples and
University of Naples "Federico II", 80125 Napoli, Italy \\
$^2$ Department of Information Engineering and INFM-Coherentia,
Second University of Naples, 81031 Aversa (CE), Italy  \\
$^3$ Department of Environmental Sciences and INFM,
        Second University of Naples, 81100 Caserta, Italy}

\maketitle

\begin{abstract}
Many natural phenomena exhibit power law behaviour in the distribution of
event size. This scaling is successfully reproduced by
Self Organized Criticality (SOC). On the other hand, 
temporal occurrence in SOC models has a Poisson-like statistics,
i.e. exponential behaviour in the inter-event time distribution, 
in contrast with experimental observations.
We present a SOC model with memory: 
events are nucleated not only as a consequence of
the instantaneous value of the local field with respect to the firing 
threshold, but on the basis of the whole history of the system. 
The model is able to reproduce the complex behaviour of inter-event time
distribution, in excellent agreement with experimental seismic data.

\end{abstract}


\maketitle

\vskip 1cm

After the pioneering work of Bak, Tang and Wiesenfeld \cite{Tang},
Self Organized Criticality (SOC) has been proposed as a successful
approach to the understanding of scaling behaviour in many natural
phenomena. The term SOC usually refers to a mechanism of slow
energy accumulation and fast energy redistribution driving the
system toward a critical state. The prototype of SOC systems is the
sand-pile model in which particles are randomly added on a two
dimensional lattice. When the number of particles $\sigma _i$ in
the $i$-th site exceeds a threshold value $\sigma _c$, this site
is considered unstable and particles are redistributed to nearest
neighbor sites. If in any of these sites $\sigma _i > \sigma _c$,
a further redistribution takes place propagating the avalanche.
Border sites are dissipatives and discharge particles outside. The
system evolves toward a critical state where the distribution of
avalanche sizes is a power law obtained without fine tuning: no
tunable parameter is present in the model.
The simplicity of the mechanism at the basis of SOC has suggested
that many physical and biological phenomena characterized by power
laws in the size distribution, represent natural realizations of
the SOC idea. For instance, SOC has been proposed to model
earthquakes \cite{Bak,Sornette},
the evolution of biological systems \cite{Sneppen}, solar flare
occurrence \cite{solar}, fluctuations in confined plasma
\cite{plasma} snow avalanches\cite{snow} and rain fall \cite{rain}.

Moreover, SOC models can be also considered as cellular automata
generating stochastic sequences of events.
An important quantity showing evidence of time correlations
in a sequence is the distribution of time intervals  between
successive events. Defining $\Delta t$ as the time elapsed between
the end of an avalanche and the starting of the next one, for the
sand-pile model one obtains that $\Delta t$ is exponentially
distributed \cite{Vulpiani}. This behaviour reveals the absence of
correlations between events typical of a Poissonian process.
Conversely  the inter-event time distribution $N(\Delta t)$ of
many physical phenomena has a non-exponential shape, as for
instance in the case of earthquakes \cite{Corral},  solar flares
\cite{Vulpiani} and confined plasma \cite{Carbone}. The failure in
the description of temporal occurrence is generally considered the
main restriction for the applicability of SOC ideas to the
description of the above phenomena.

In this letter we address the problem of introducing time correlations
within SOC and, in order to validate our model, we compare our results
with experimental records from seismic catalogs.
Seismicity is considered here as a typical physical process with
power law in the size distribution but also strong correlations
between events. In this case, the sand pile model can
be directly mapped \cite{Ito,Brown,Olami} in the Burridge-Knopoff
model, proposed for the description of earthquake occurrence. In
this model a continental plate is represented as a series of
blocks interconnected with each other and with a rigid driver
plate by springs, then the quantity $\sigma _i$ represents the
global force acting on the $i$-th block and $\sigma _c$ the
threshold for slippage. We introduce memory within the SOC context: 
the local instability depends not only on
the instantaneous value $\sigma _i$ but on the whole history of energy
accumulation. Our memory ingredient is analogous
to recent ideas \cite{Stein} introduced for
the understanding of earthquake interactions.

The first observation of correlations between earthquakes, dates
back to Omori \cite{Omori} who suggested that earthquakes
tend to occur in clusters temporally located after main events:
the number of aftershocks following a main event after a time $t$,
$r(t)$, decays as a power law $r(t) \sim 1/t$. Furthermore, a large
earthquake produces also an abrupt
 modification in seismic activity across a widespread area \cite{Hill}. A
 striking example of this remote triggering mechanism is the Landers
earthquake of magnitude 7.3 occurred in 1992, which triggered three
 hours later the 6.5 magnitude event in the town of Big Bear on a
 different fault, together with a general increase in activity across
much of the Western United States. A clear understanding of the
 physical processes responsible for this behaviour is still lacking:
the non-local triggering mechanism cannot be explained in terms of
static stress changes responsible for aftershocks and 
one must invoke non linear interactions to 
modify the friction law on remote faults \cite{Hill,Dietriech,nota,Stein}.

 The inter-time distribution combines both the  effect of the local
 clustering of the aftershocks sequence described by the Omori law,
 with the remote triggering mechanism involving larger distances.
 The presence of both features give rise to an intertime distribution
$N(\Delta t)$ that is not a power law but has a more complex shape
\cite{Mega}. Nevertheless, Corral has shown that this shape is
quite independent on the geographical region and the magnitude
range considered \cite{Corral}. This observation indicates that
$N(\Delta t)$ is a fundamental quantity to characterize the
temporal distributions of earthquakes.

Here we introduce within SOC a non local mechanism for
event nucleation. In our approach seismic fracture depends on
a collective behaviour of the earth crust: the triggering of a new
event is determined by the combined effect of the increase in the
static stress together with the local weakening in a fault due to
the loading global history. To this extent, we consider a square
lattice of size $L$, each site being characterized not only by the
value of the local stress $\sigma _i$ but also by a site-counter
$c_i$ that represents the local memory. At $t=0$ local stresses
are assigned at random between $\sigma _c -z$ and $\sigma _c$,
where  $z$ is the lattice coordination number and $\sigma _c>z$,
whereas $c_i$ is randomly set between zero and one.
The simulation proceeds as follows. At each time $t$ all
sites are loaded with an uniformly increasing external stress, by
adding one unit to all $\sigma _i$'s, and the local variable $p_i$
is defined as
\be
p_i={(\sigma _i-\sigma _c+z) \over z} \qquad \qquad {\rm if }
\quad   \sigma _c-z \leq \sigma _i \leq \sigma _c
\label{ggg}
\ee
whereas $p_i=1$   if  $\sigma _i>\sigma_c$  and $p_i=0$ if $\sigma
_i < \sigma _c-z$.
Then at each site the quantity $\alpha _i = {(1-c_i) \over p_i}$
is evaluated, measuring the local instability with respect to
slippage, and its minimum value in the system, $\alpha _{min}$, is
found.  This value indicates the site most susceptible to seismic
failure because of both the high local stress at that instant of
time and the cumulated history of loads saved into the counters
$c_i$. If $\alpha _{min}$ is larger than a critical value $\alpha _c$,
all counters are updated as
\begin{equation}
c_i^{new}= c_i^{old} +\alpha _{min} p_i
\label{uno}
\end{equation}
Then the external stress is uniformly loaded at constant rate
and at each step the new value
of $\alpha_{min}$ in the system is evaluated and Eq.(2) applied.
As soon as $\alpha_{min} < \alpha_c$, the site $i$ with
$\alpha_i = \alpha_{min}$ becomes
the epicenter where the earthquake nucleates and its counter is set to zero.
Other sites with
$\alpha_{min} < \alpha_i < \alpha_c$ are considered stable unless involved
in the fracture propagation. This choice is expression of fracture being a
phenomenon controlled by the extreme value statistics.

When a site nucleates an earthquake, it discharges elastic energy
uniformly by the transfer of a unit stress to all nearest
neighbours, as in the sand pile model.
The process goes on letting unstable sites, characterized by
$\alpha_i < \alpha_c$,
discharge energy and in this way propagating the seismic event
farther and farther from the epicenter.
The counters of all discharging sites are set to zero during the evolution,
whereas counters of all other sites are updated at the end
according to  Eq.(2) with the actual value of $\alpha_{min}$.
Energy back-flow allows to activate sites found stable during the forward
propagation triggering further energy redistributions.
At the end of the process the external load is
increased again at constant rate until another event takes place.

The updating rule (2) is equivalent to consider a time
dependent friction law \cite{Dietriech}, whose evolution is
controlled not only by the local state at
 previous times but also by the instability condition for the whole
 system.  Eq.(\ref{uno}) then introduces long range interactions and remote
triggering by means of  $\alpha_{min}$, since all sites in the
system, even far from the epicenter, share this common
information. The more a site is stressed (high $p_i$), the
stronger it will react to this information.
We have checked that our results are
substantially unchanged if Eq.(2) is applied to a finite region of size $l<L$ 
centered in the site with $\alpha_i =\alpha_{min}$, for large enough $l$. 
A breaking rule similar to Eq. (2)
has been successful in providing a good description of  the
propagation of stress corrosion cracks \cite{Herrmann}, a fracture process
where local mechanical resistance of materials is weakened
in time by chemical agents.

It is possible to calculate in a mean field approximation
the fraction of active sites as function of time.
This approximation is based on the assumption that
the external stress is kept fixed and therefore can
describe the behaviour of the system only at short time scales.
As a consequence of this hypothesis, the fraction of active sites is related
to the rate of occurrence of aftershocks, $r(t)$, happening over time
scales shorter than the characteristic time of the loading mechanism.
Let us consider
at each time $t$ and each site the quantity $q_i(t)=1-\alpha_i$ for
$0\le \alpha_i \le 1$ and $q_i(t)=0$ for $\alpha_i >1$.
According to Eq.(\ref{uno})
one has, at constant load condition, that the value of $q_i(t)$ 
at the next time step is given by
$q_i(t+1)=q_i(t)+\alpha_{min}$, if the $i$-th site does
not discharge energy and $q_i(t+1)=0$ otherwise.
Hence, the statistical average is
\be
\langle q_i(t+1)\rangle=\big ( \langle q_i(t)\rangle+\langle \alpha_{min} 
\rangle \big ) P^s_i(t)
\label{mf}
\ee
 with $P_i^s(t)$ the probability for the site $i$ to be
stable at time $t$. Since, the spatial
average, $q(t)={1\over L^2}\sum_{i=1}^{L^2} q_i(t)$,
is an estimate of the fraction of sites that could become active at the
next time step, $q(t)$ is the
probability for a generic site to be unstable at time $t$ and then
$q(t) \simeq 1-P^s_i(t)$.
In the hypothesis $ \alpha_{min} \ll \alpha_{i}$, valid for a
system close to trigger an event, one can
neglect $\alpha_{min}$ in Eq.(\ref{mf}). Finally, supposing that $q_i(t)$
 is a self-averaging quantity  one has
$q(t+1)  \simeq q(t) \big (1- q(t)  \big )$, which gives the Omori
law $q(t) \sim t^{-1}$.

In order to evaluate the complete inter-time distribution $N(\Delta
t)$, we numerically generate a large statistics of events and calculate
the time distance between every couple of successive events involving
more than one single site.
 Fig.1 shows, the experimental and the numerical data for
$\alpha_c=0.9$ and $L=500$. By re-scaling the
 numerical waiting time with an appropriate constant value,
our data provide a very good agreement with the experimental
distribution from the Southern California Catalogue \cite{Cali},
 whereas the original SOC model exhibits exponential decay.
We have monitored the  behaviour of the distribution for different
values of $\alpha_c$ (Fig.2): 
For $\alpha_c\le 0.3$, an exponential decay is observed whereas for 
intermediate values of $\alpha_c$ a complex behaviour starts to set in.
For $\alpha_c \ge 0.7$ the data follow a unique universal curve.

In order to fully validate the theoretical ideas of our model,
we have also analyzed other statistical properties of seismic
catalogues:
 energy and epicenters distance distributions.
 The energy release in an earthquake is expressed in terms of the
 magnitude, which is proportional to the logarithm of the fractured 
area $A$ 
\begin{equation}
M=k Log(A)+M_{min} 
\label{magni}
\end{equation}
where $M_{min}$ is a constant depending on the area units.
The Gutenberg-Richter law implies that $N(M)$, the number of earthquakes
with magnitude $M$, follows an exponential law \cite{Gutenberg}
\begin{equation}
N(M) \sim 10^{-b M}
\label{GR}
\end{equation}
where $b$ is an experimental constant generally close to one.
In order to compare our numerical results with experimental data, 
we evaluate the magnitude of an event using
Eq. (\ref{magni}) with $A$ being the number of discharging sites and
$k \simeq 0.65$ as for the Southern California Catalogue.
The choice of the minimum magnitude $M_{min}$ is arbitrary
since it is related to the unit cell area, and in our case we set
$M_{min}=2$.
The magnitude distribution, after an initial transient at small
magnitudes $(M<3)$, follows the expected exponential behaviour of the
Gutenberg-Richter law (Fig.3) over a magnitude range increasing with
the system size $L$. Strong fluctuations observed at large $M$ for small
system sizes are finite size effects.
The value of the best fit exponent $b$ depends on the parameter $\alpha_c$
and becomes parameter independent  for $\alpha_c \ge 0.7$, where $b \sim 0.84$
(inset Fig.3). Comparing our numerical results with the data from the
California Catalogue
(Fig.4) good agreement is found between the experimental
best fit value $b_{exp} \sim 0.86$ and numerical prediction. In
the non conservative case, SOC also provides good
agreement with the experimental size distribution \cite{Olami}.

Seismic catalogues also record the spatial coordinates of earthquake
 epicentres. We numerically evaluate the
 cumulative distribution of distance between all possible couple
 of events at a distance smaller than $d$, $N(d)$.
We obtain a power law behaviour with an
exponent equal to 1.84. Agreement with experimental data is observed for
small distances. $N(d)$
calculated for the original SOC model  provides similar results.

It is worth noticing that the behaviour of all experimental
 distributions is reproduced by numerical simulations without
any fine tuning, i.e. numerical results are parameter independent
for $\alpha_c>0.7$.


The complex seismic activity over large regions of the world is
 controlled by both local stress redistributions, generating
 aftershocks, and long range load transfer in the surrounding crust.
 In our approach these mechanisms are simply implemented in
self-consistent local laws containing long range memory of stress
 history. A large event could then increase the seismic activity by
inducing global weakening in the system, or else could even inhibit
 future earthquakes by resetting the local memory.
 This global memory ingredient could correspond to a variety of
 physical mechanisms inducing weakening in time for real
faults \cite{Benn},
as  stress corrosion \cite{Herrmann}, fault gauge deterioration
\cite{Mora} or pore pressure variation \cite{Mulargia}.
  
We have considered earthquake triggering as an example of physical
problems in which time correlations are extremely important.
We suggest that this SOC model with memory may be
relevant for other physical
phenomena described by a SOC approach and exhibiting a
non-exponential decay in the inter-time distribution \cite{Vulpiani}.

{\small Acknowledgements. We would like to thank A. Coniglio,
 P. Gasparini and S. Nielsen for helpfull discussions.
This work is part of the project of the
Regional Center of Competence "Analysis and Monitoring of
Environmental
Risk" supported by the European Community on Provision 3.16.}

\begin{figure}
\includegraphics[width=12cm]{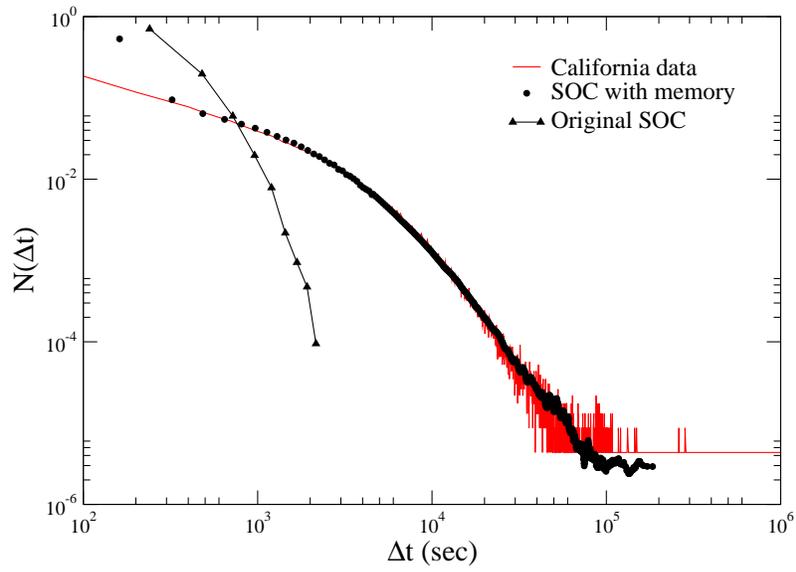}
\caption{ (Color online) The normalized  waiting time distribution 
$N(\Delta t)$
between successive events for experimental and numerical data for
our model and the original SOC model. Data for our model correspond
to 2500 events in 3000 configurations of system size $L=500$ with
$\alpha_c=0.9$.
Time is measured in seconds and all numerical waiting times are rescaled 
by the constant factor 190.}
 \label{fig1}
\end{figure}

\begin{figure}
\includegraphics[width=12cm]{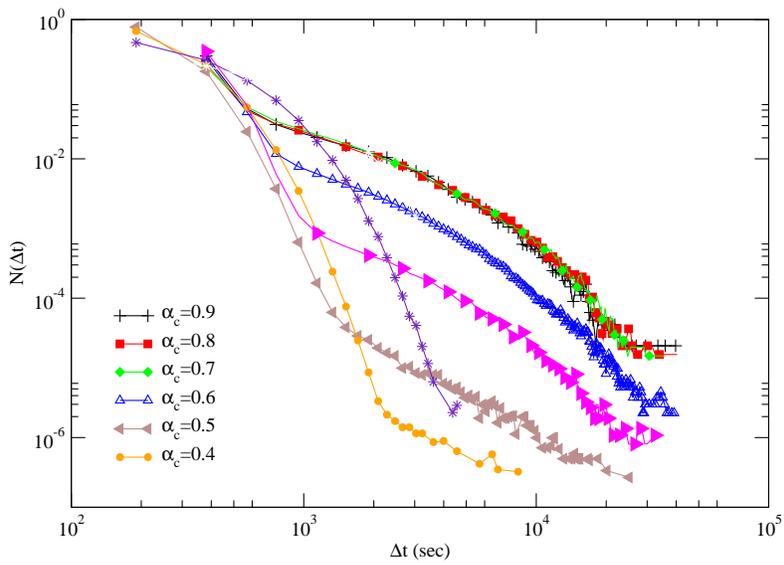}
\caption{(Color online) Inter-event time distribution for 2500 events in
1000 configurations of system size $L=100$ and 
different values of $\alpha_c$.}
\end{figure}

\begin{figure}
\includegraphics[width=12cm]{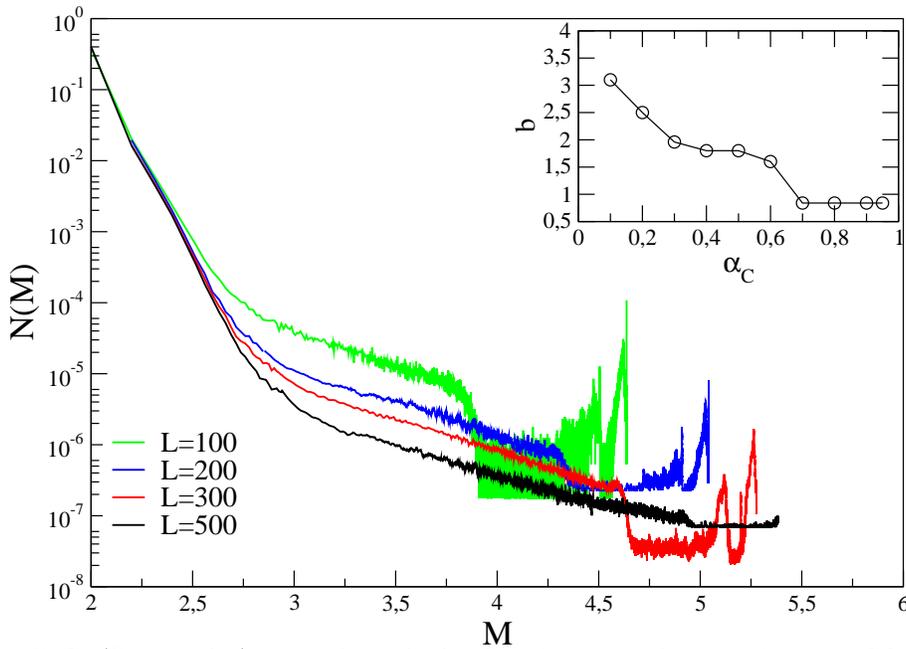}
\caption{(Color online) Normalized distribution of the magnitude $M$ 
for $\alpha_c=0.9$ and increasing lattice sizes $L$ from top to bottom. The
horizontal cutoff is due to single event statistics. In the inset
we show the best fit exponent $b$ of the Gutenberg-Richter law,
Eq.(\ref{GR}), as a function of  $\alpha _c$.
}
\end{figure}

\begin{figure}
\includegraphics[width=12cm]{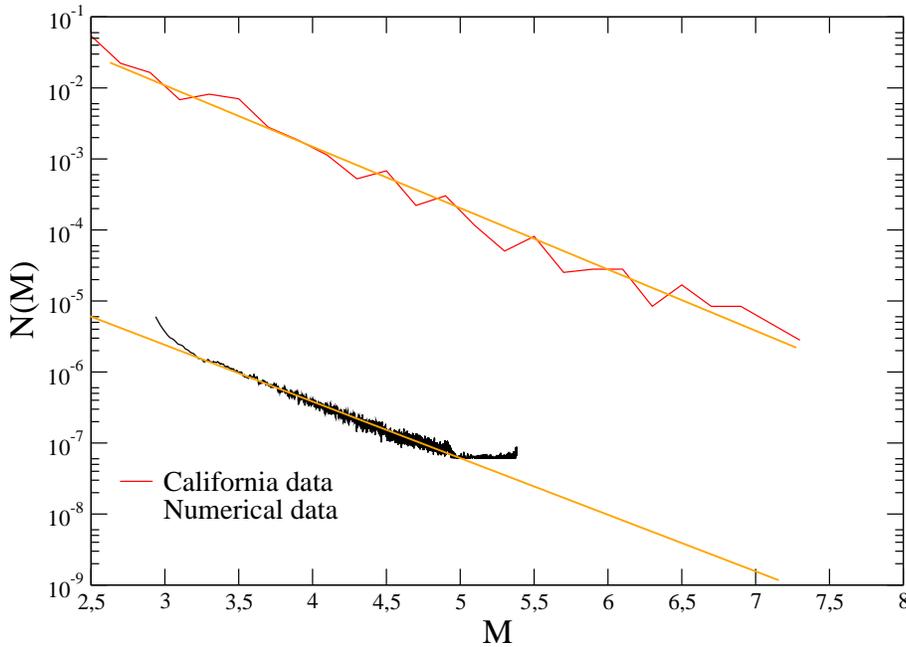}
\caption{(Color online) The comparison of $N(M)$ from experimental (top) 
and numerical
data (bottom) for $L=500$ and  $\alpha _c =0.9$. The average is taken over
3000 configurations of 2500 events each. Straight lines are the
best fit curves $10^{-0.86 M}$ and $10^{-0.84 M}$ respectively.
}
\end{figure}


\begin{thebibliography}{}
\bibitem{nota} In fact, these
variations decrease with distance $r$ from the source as $r^{-3}$
and, for instance, in the case of the Landers earthquake the
measured static stress change at a distance of 300 Km was less
than $10^{-2}$ bar.
\end{thebibliography}
\end{document}